\documentclass[aps,prb,twocolumn,superscriptaddress,showpacs]{revtex4}
\usepackage{graphicx}
\usepackage{latexsym}
\usepackage{amssymb}
\usepackage{amsmath}
\usepackage{amsfonts}
\usepackage{bm}
\usepackage{multirow}
\usepackage{color}
\usepackage{comment}
\newcommand{\ii}{\mathrm{i}}

\newcommand{\U}{\mathrm{U}}

\newcommand{\beq}{\begin{equation}}
\newcommand{\eeq}{\end{equation}}
\newcommand{\beqn}{\begin{eqnarray}}
\newcommand{\eeqn}{\end{eqnarray}}

\DeclareMathAlphabet{\mathbbold}{U}{bbold}{m}{n}

\def\U{{\rm U}}

\begin{document}

\title{Interacting Valley Chern Insulator and its Topological Imprint on Moir\'{e} Superconductors}


\author{Xiao-Chuan Wu}
\affiliation{Department of Physics, University of California,
Santa Barbara, CA 93106, USA}

\author{Yichen Xu}
\affiliation{Department of Physics, University of California,
Santa Barbara, CA 93106, USA}

\author{Chao-Ming Jian}
\affiliation{Kavli Institute of Theoretical Physics, Santa
Barbara, CA 93106, USA}

\author{Cenke Xu}
\affiliation{Department of Physics, University of California,
Santa Barbara, CA 93106, USA}

\begin{abstract}

One salient feature of systems with Moir\'{e} superlattice is
that, Chern number of ``minibands" originating from each valley of
the original graphene Brillouin zone becomes a well-defined
quantized number because the miniband from each valley can be
isolated from the rest of the spectrum due to the Moir\'{e}
potential. Then a Moir\'{e} system with a well-defined valley
Chern number can become a nonchiral topological insulator with
$\U(1) \times Z_3 $ symmetry and a $\mathbb{Z}$ classification at
the free fermion level. Here we demonstrate that the strongly
interacting nature of the Moir\'{e} system reduces the
classification of the valley Chern insulator from $\mathbb{Z}$ to
$\mathbb{Z}_3$, and it is topologically equivalent to a bosonic
symmetry protected topological state made of local boson
operators. We also demonstrate that, even if the system becomes a
superconductor when doped away from the valley Chern insulator,
the valley Chern insulator still leaves a topological imprint as
the localized Majorana fermion zero mode in certain geometric
configuration.

\end{abstract}

\maketitle

\section{Introduction}

Systems with Moir\'{e} superlattice not only have demonstrated
amazing correlated physics such as correlated insulator at
fractional filling and also
superconductivity~\cite{wangmoire,mag01,mag02,young2018,TLGSC,FM,kimtalk,efetov,TDBG1,TDBG2,TDBG3},
due to the narrow width of the Moir\'{e}
minibands~\cite{flat1,flat2,flat3,flat4,TDBG1,TDBG2,TDBG3}, recent
theoretical and experimental studies have shown that under certain
conditions some of the Moir\'{e} systems can have novel
topological features. The reason is that, within the mini
Brillouin zone, the miniband that originates from each valley of
the original graphene Brillouin zone can be isolated from the rest
of the spectrum, which makes the Chern number of a miniband from
each valley a well-defined quantized Chern number. Indeed, many
theoretical works have predicted that the Chern number of miniband
from each valley can take different integer values in various
Moir\'{e} systems, depending on the displacement field and also
the twisted angle~\cite{band3,band5,TDBGt,dai2}. Recently
quantized~\cite{hall} (or unquantized~\cite{FM}) Hall conductivity
has been observed at doping with one or three electrons/holes per
Moir\'{e} unit cell away from charge neutrality in some Moir\'{e}
systems, whose most natural explanation is that under the
experimental conditions the Moir\'{e} system forms a fully spin
and valley polarized state~\cite{zaletel,zhangmao,hall}, hence a
miniband with a net Chern number is completely filled.
Although superconductivity has not been observed near the
insulating states with nonzero Hall conductivity, these
topological physics is certainly interesting in its own right.

Here we study the non-chiral valley Chern insulator (VCI), namely
the valley space is not polarized, while instead we assume the
spin space is fully polarized. These assumptions may be relevant
to the twisted double bilayer graphene (TDBG) at half-filling (two
extra electrons per Moir\'{e} unit cell) away from charge
neutrality~\cite{kimtalk,TDBG1,TDBG2,TDBG3}, for the following
reasons:

{\it 1.} Calculations with various methods have demonstrated that
in TDBG the minibands from each valley~\cite{band3,TDBGt,dai2}
could have nonzero quantized Chern number depending on the
displacement field and also twisted angle. Symmetry of the system
guarantees that the degenerate minibands from the two valleys must
have opposite Chern numbers. If the partially filled miniband has
zero valley Chern number (with certain displacement field), then
the formalism developed in
Ref.~\onlinecite{xuleon,kivelson,xuFM,fumag} becomes applicable.
In this work we assume a nonzero valley Chern number of the
minibands.

{\it 2.} The charge gap of the insulator at half-filling increases
with inplane magnetic field~\cite{kimtalk,TDBG1,TDBG2,TDBG3},
whose main effect is most likely a Zeeman coupling with spin. This
phenomenon suggests that the insulator at half-filling observed in
this system has a fully polarized ferromagnetic order.

{\it 3.} superconductivity is observed near half-filling, which
suggests that the valley space is likely unpolarized, because
otherwise electron states with momentum $\vec{k}$ and $-\vec{k}$
would be nondegenerate, which makes pairing with total zero
momentum difficult. Even within each valley, the $C_3$ symmetry of
the Fermi surface still does not guarantee degeneracy of states
with momentum $\vec{q}$ and $-\vec{q}$, where $\vec{q}$ is the
momentum away from each valley.

These observations, plus simple counting suggest that at
half-filling (with two extra electrons per Moir\'{e} unit cell
away from the charge neutrality), the insulator observed in TDBG
could be a spin-polarized ``band insulator" where spin down
electrons fill both minibands whose total valley Chern number is
zero for relatively weak displacement field~\cite{dai2}, while
spin-up electrons only fill the valence miniband with nonzero
Chern number. Then the system becomes a spin polarized nonchiral
topological insulator, or a valley Chern insulator (VCI). The
symmetry of this VCI is $\U(1) \times Z_3 $, where the $\U(1)$
corresponds to the charge conservation, $Z_3$ is the conservation
of the ``valley quantum number" (valley momentum of the original
graphene Brillouin zone is conserved mod 3).

\section{Interacting Valley Chern Insulator}

At the free fermion level, the Hamiltonian of the $1d$ edge state
of the VCI with valley Chern number $C = 1$ is ($C = 1$ means the
minibands from the right and left valleys have Chern number $\pm
1$ respectively) \beqn H = \int dx \ \psi^\dagger \left( - \ii
\tau^z
\partial_x \right) \psi, \label{edge} \eeqn $\tau^z = \pm 1$
represents the edge modes from the right and left valleys. The
$Z_3$ symmetry acts on the boundary electron operator as $Z_3:
\psi \rightarrow \left( \exp(\ii \frac{2\pi}{3} \tau^z) \right)
\psi$. the $Z_3$ symmetry guarantees that no fermion bilinear mass
term can be added to the boundary Hamiltonian. Also, for arbitrary
copies of the VCI, or for states with arbitrary valley Chern
number, fermion bilinear mass operators are always forbidden by
the $Z_3$ symmetry. Hence the classification of VCI in the
noninteracting limit is $\mathbb{Z}$.

The Moir\'{e} systems are intrinsically strongly interacting
systems, thus we need to understand the effects of interaction on
the VCI. In the last decade, examples have been found where
interaction can indeed change the classification of topological
insulators~\cite{fidkowski1,fidkowski2,chenhe3B,senthilhe3,youinversion,qiz8,zhangz8,yaoz8,levinguz8},
thought usually interaction would reduce the classification of a
topological insulator from $\mathbb{Z}$ to $\mathbb{Z}_8$, or
$\mathbb{Z}_{16}$. In the following we will demonstrate that an
interacting VCI actually has a $\mathbb{Z}_3$ classification.

The VCI can be naturally embedded into a nonchiral topological
insulator (TI) with $\U(1)_c \times \U(1)_s$ symmetry, which is
analogous to the quantum spin Hall insulator, and the valley space
can be viewed as a pseudo-spin space, with pseudospin $\tau^z =
\pm 1$ labelling the right and left valley spaces. We will start
with the TI with $\U(1)_c \times \U(1)_s$ symmetry~\footnote{more
precisely the symmetry of this system is in fact $\frac{\U(1)_c
\times \U(1)_s}{Z_2}$, which means that the $Z_2$ subgroup of
$\U(1)_c$ is identified with that of the $\U(1)_s$, but we will
ignore this subtlety.}. There are only two elementary fermions
with charge $(1,1)$ and $(1,-1)$ under the $\U(1)_c\times \U(1)_s$
symmetry, and for the simplest case they form Chern insulators
with Chern number $\pm 1$ respectively. Such a TI has a
topological response:
\begin{align}
\mathcal{L}^{r} = \frac{2\ii}{2\pi} A_c \wedge d A_s,
\label{Eq:QSH_response}
\end{align}
where $A_c$ and $A_s$ are background gauge fields that couple to
the $\U(1)_c$ and $\U(1)_s$ symmetries, and an electron carries
charge-1 under $A_c$ and charge $\pm 1$ under $A_s$.

In the following, we will consider the consequence of breaking
$\U(1)_s$ to its discrete subgroups. When we break $\U(1)_c \times
\U(1)_s$ symmetry down to $\U(1)_c \times \mathbb{Z}_n$, the
fermions transform under $Z_n$ as
\begin{align}
Z_n:  \psi_1 \rightarrow e^{2\pi \ii /n}\psi_1,~~~\psi_2
\rightarrow e^{-2\pi \ii /n}\psi_2.
\end{align}
To describe the nonchiral TI, we can use the $K-$matrix
formalism~\cite{wenzee}. The system can be described by the
following Chern-Simons theory \beqn \mathcal{L} = \frac{\ii}{4\pi}
\sum_{A,B=1,2} K^{AB} a^A \wedge d a^B, \ \ \ K = \left(
\begin{array}{cc}
1 & 0 \\
0 & -1
\end{array}
\right), \eeqn where $a^A$ with $A=1,2$ are two dynamical U(1)
gauge fields. The edge state of this TI is described by the
Luttinger liquid theory with two chiral boson fields $\phi_{1}$,
$\phi_2$ and the same $K-$matrix above~\cite{wenedge,wenreview}:
\begin{align}
\mathcal{L}_{\rm edge} = \sum_{A,B=1,2}\frac{K^{AB}}{4\pi}
\partial_x \phi_A \partial_t \phi_B - \frac{V^{AB}}{4\pi}
\partial_x \phi_A \partial_x \phi_B,
\label{Eq:Edge_Luttinger_Luquid}
\end{align}
where $V$ is a $2\times 2$ positive-definite velocity matrix. In
this theory, the boson fields satisfy the equal time commutation
relation $[\phi_A(x), \partial_y \phi_B(y)]= 2\pi \ii
(K^{-1})^{AB} \delta(x-y)$. Under the $\U(1)_c \times Z_n$
symmetry, the chiral boson fields $\phi_{1,2}$ transform as \beqn
\U(1)_c &:&  \phi_{1,2} \rightarrow \phi_{1,2} + \alpha, \ \ Z_n :
\phi_{1,2} \rightarrow \phi_{1,2} \pm \frac{2\pi}{n}. \eeqn

Now we demonstrate that the nonchiral TI with $\U(1)_c \times Z_n$
symmetry at most has a $\mathbb{Z}_n$ classification under local
interaction, for odd integer $n$. The fact that $n$ copies of such
TI together are topologically trivial can be seen from the edge
theory of this system which consists of $n$ copies of Luttinger
liquid theory Eq. \ref{Eq:Edge_Luttinger_Luquid}. Let's denote
chiral boson fields in this $n$-copy Luttinger liquid theory as
$\phi_{i,A}$ where $i=1$ is the copy index and $A=1,2$ is the
label for the chiral bosons within each copy. The boundary of $n$
copies of the TI can be gapped out by the following symmetric
boundary interaction without ground state degeneracy: \beqn
\mathcal{L}_{\rm edge}^{(1)} &=& - g \cos\left( \sum_{i = 1}^n
(\phi_{i,1} - \phi_{i,2}) \right) \cr\cr &-& g' \sum_{i = 1}^{n-1}
\cos(\phi_{i,1} + \phi_{i,2} - \phi_{i+1,1} - \phi_{i+1,2}).
\label{edge}\eeqn These are local interacting terms between the
electrons.


This edge theory $\mathcal{L}_{\rm edge}^{(1)}$ can be analyzed
systematically as following: There are in total $n$ different
terms in $\mathcal{L}_{\rm edge}^{(1)}$, and we can represent each
term in $\mathcal{L}_{\rm edge}^{(1)}$ as $\cos(\mathbf{\Lambda}_I
\cdot \mathbf{\Phi})$. $\mathbf{\Lambda}_I $ are $2n$ component
vectors ($I = 1, \cdots n$), and $\mathbf{\Phi} = (\phi_{1,1},
\phi_{1,2}, \phi_{2,1} \cdots)$. $\mathbf{\Lambda}_I$ are a set of
minimal linearly independent integer vectors, and they satisfy the
condition~\cite{nullhaldane,nulllevin} \beqn \mathbf{\Lambda}^t_I
\mathbf{K}^{-1} \mathbf{\Lambda}_J = 0, \label{null} \eeqn for any
$I, J = 1, \cdots n$. Here $\mathbf{K}$ is the $2n \times 2n$
block-diagonal $K-$matrix for $n$ copies of the nonchiral TI with
$\U(1)_c \times Z_n$ symmetry. Eq.~\ref{null} implies that the
arguments in all the cosine terms in $\mathcal{L}_{\rm
edge}^{(1)}$ commute with each other, and hence all terms in
$\mathcal{L}_{\rm edge}^{(1)}$ can be minimized
simultaneously.

The $2n-$component integer vector $\mathbf{\Lambda}_I$ is a vector
in a $2n$ dimensional cubic lattice with lattice constant $1$.
Linear combinations of $\mathbf{\Lambda}_I$ span a $n-$dimensional
hyperplane of this $2n$ dimensional cubic lattice. Linear
combinations of $\Lambda_I$ span a $n-$dimensional hyperplane of
this $2n$ dimensional cubic lattice. To be rigorous we also need
to show that $\Lambda_I$ are the irreducible basis vectors of the
lattice sites residing on this $n-$dimensional hyperplane, hence
the minimum of $\mathcal{L}_{\rm edge}^{(1)}$ has no degeneracy.
This has been verified for odd integer $n$, which leads to a fully
gapped edge state. Similar reduction of classification was also
discussed in a different context~\cite{ashvinzn}.


The formalism above will lead to the same conclusion with other
observations. For example, the response to the external gauge
field Eq.~\ref{Eq:QSH_response} implies that for a $C = 3$ VCI, a
$2\pi$ flux of the electromagnetic field would nucleate a trivial
$Z_3$ charge according to Eq.~\ref{Eq:QSH_response}. The fact that
the VCI is generically trivial when $C = 3$ implies that this
interacting VCI can be adiabatically deformed into a spatially
direct product state under interaction without closing the gap in
the bulk, and the deformation process preserves all the
symmetries. This result is relevant to the TDBG with certain
displacement field~\cite{TDBGt,dai2}, and the valence/hole
miniband of the heterostructure of trilayer graphene and hexagonal
boron nitride~\cite{band3,band5}. In these cases calculations
suggest that the valley Chern number is $C = 3$ in the
noninteracting limit.

There is a three-body interaction (six-body term) in
Eq.~\ref{edge} that breaks the valley-$\U(1)$ symmetry (analogue
of the $\U(1)_s$ in quantum spin Hall insulator) to the physical
$Z_3$ symmetry. This three-body interaction originates from a
second order effect of lattice scale short-range interactions
which allow large momentum transfer. Here we will give a
rudimentary estimate of the strength of this effect. We consider a
scattering process of three particle wave packets from valley$-1$
to valley$-2$: $\{ \vec{Q}, \vec{Q}, \vec{Q} \} \rightarrow \{ -
\vec{Q}, - \vec{Q}, - \vec{Q} \}$, where $\vec{Q}$ is the momentum
of valley$-1$ in the original graphene Brillouin zone. Each wave
packet has the size of the Moir\'{e} unit cell $l^2$, where $l$ is
the Moir\'{e} lattice constant and $l \sim 1/(2 \sin(\theta/2))$
(we are viewing $l$ as the dimensionless ratio between the Moire
lattice spacing and the lattice constant of the original
graphene). This process can come from a second order effect of two
combined processes: $\{ \vec{Q}, \vec{Q}\} \rightarrow \{ -
\vec{Q}, \Gamma \}$, and $\{ \vec{Q}, \Gamma \} \rightarrow \{ -
\vec{Q}, - \vec{Q}\}$, where $\Gamma$ corresponds to the state
with momentum 0 in the graphene Brillouin zone. The strength of
this scattering process is roughly $g \sim l^2 U^2/(E_\Gamma)
1/l^6$, where $U$ is the total short-range interaction in each
unit cell of the original lattice. For example, since we are
mostly considering spinless fermions motivated from the fact that
in TDBG the electron spins are fully polarized in a large region
of the phase diagram, we only consider the nearest neighbor
repulsions, which is about $5.5$eV according to
Ref.~\onlinecite{hubbard}. There are six nearest neighbor pairs of
sites (in total eight sites) within one unit cell of the double
bilayer graphene, including both inplane and vertical pairs. Hence
we roughly estimate $U \sim 5.5 \times 6$eV. $E_\Gamma$ is the
energy of the electron at momentum 0, which is about $4.9$eV
choosing the same parameter as Fig.3 of Ref.~\onlinecite{RMP}. The
first factor $l^2$ is from summing over all sites in the Moir\'{e}
unit cell, and the last factor $1/l^6$ is due to the fact that, if
the wave function packet is normalized in each Moir\'{e} unit
cell, the amplitude of the wave function on each site will be
$1/l$. Choosing $\theta = 1.33$ as in experiment
Ref.~\onlinecite{kimtalk,TDBG2}, we eventually obtain $g \sim
0.65K$, which is not strong, but its effect is still observable
experimentally. If further neighbor interaction in graphene is
considered~\cite{hubbard}, this effect will be further enhanced.

The valley conservation can also be broken at the single-particle
level due to inter-valley scattering from the Moir\'{e} potential.
But since the Moir\'{e} potential is naturally a smooth function
of space, this inter-valley tunnelling (which involves a large
momentum transfer) is expected to be suppressed exponentially with
the unit cell size $l$, while the estimated three-body interaction
decays as a power-law of $l$. Hence with small angle, the
three-body interaction actually dominates the single particle
inter-valley tunnelling. Hence with large $l$ (small twisting
angle $\theta$) we can safely assume that the valley conservation
is conserved at the single particle level, but broken down to
$Z_3$ by interaction.

\section{Bosonic symmetry protected topological state}

For odd integer $n$, a general connection between the nonchiral TI
with $\U(1) \times Z_n $ symmetry and bosonic symmetry protected
topological (bSPT) state~\cite{wenspt,wenspt2} can be made. Since
the interacting TI has a $\mathbb{Z}_n$ classification, one copy
of the elementary TI is topologically equivalent to $n+1$ (an even
integer) copies of the TI; while according to
Ref.~\onlinecite{xugraphene,spn,xufb,hermelecrystal}, even number
of such TIs can be ``glued" into a bSPT state with the same
symmetry under interaction, where all the local fermion
excitations at the boundary are gapped out by interaction, leaving
only symmetry protected gapless local bosonic excitations.

A variety of bSPT states and their edge states can be described by
the Chern-Simons theory with the following
$K$-matrix~\cite{luashvin}, whose boundary state is described by
two chiral bosons $\varphi$ and $\theta$ with $K-$matrix:
\begin{align}
K_{\rm bSPT} = \left(
\begin{array}{cc}
0 & 1 \\
1 & 0
\end{array}
 \right).
 \label{Eq:K_bSPT}
\end{align}
The chiral bosons transform under the symmetries as \beqn \U(1)_c
&:& \varphi \rightarrow \varphi + 2\alpha, \ \ \ \theta
\rightarrow \theta \nonumber \cr\cr Z_n &:& \varphi \rightarrow
\varphi, \ \ \ \theta \rightarrow \theta - 2\pi /n. \label{Eq:
symmetry_action_bSPT_edge} \eeqn Now, we consider an $(1+1)d$
interface between the bSPT and the nonchiral TI discussed
previously which can be described by the four boson fields
$\phi_{1,2}$, $\varphi$ and $\theta$. The symmetry allowed
interaction that can gap out this interface without degeneracy is
\beqn \mathcal{L}_{\rm edge}^{(2)} \sim - u_1 \cos(\phi_1 + \phi_2
- \varphi) - u_2 \cos(\phi_1 - \phi_2 + 2\theta).
\label{Eq:QSH-BSPT_GappedInterface} \eeqn Again the arguments in
the cosine terms commute with each other, hence all terms in
$\mathcal{L}_{\rm edge}^{(2)}$ can be minimized simultaneously,
and the interface is gapped out without degeneracy through the
same reasoning as the last section. The existence of such a gapped
interface between the bSPT and the fermionic TI guarantees the
topological equivalence between the two sides of this interface.

The physical interpretation of the bosonoic fields $\varphi$ and
$\theta$ can be understood in terms of their quantum numbers. The
local boson field $e^{\ii \varphi}$ can be identified with the
bound state $\psi_1 \psi_2$;
the quantum number of the single boson operator $e^{\ii \theta}$
is equivalent to $(n+1)/2$ copies of the particle-hole pair
$\psi_1^\dag\psi_2$ ($Z_n$ charge is defined mod $n$). We can see
that {\it when and only when} $n$ is an odd integer, $e^{\ii
\theta}$ can be viewed as a local boson field. Hence for odd
integer $n$, a nonchiral TI with $\U(1) \times Z_n $ symmetry is
equivalent to a bSPT constructed with local bosons. This result
implies that, under interaction any VCI in the Moir\'{e} system
can have fully gapped single electron excitation, but meanwhile
symmetry protected gapless local boson excitations at its
boundary. This was thought to be only possible for {\it even}
copies of nonchiral TI such as the quantum spin Hall insulator
with spin $S^z$
conservation~\cite{xugraphene,spn,xufb,hermelecrystal}.

In the spin-polarized Moir\'{e} system there is one more natural
symmetry, an effective time-reversal symmetry which is a product
between the ordinary time-reversal of electron and spin flipping.
This time-reversal symmetry acts on the electron as \beqn
\mathcal{T}: \psi \rightarrow \tau^x \psi , \eeqn and it has
$\mathcal{T}^2 = +1$. All our conclusions above including the
classification of interacting VCI and its connection to BSPT still
hold with this extra effective time-reversal.

\begin{figure}
\includegraphics[width=\linewidth]{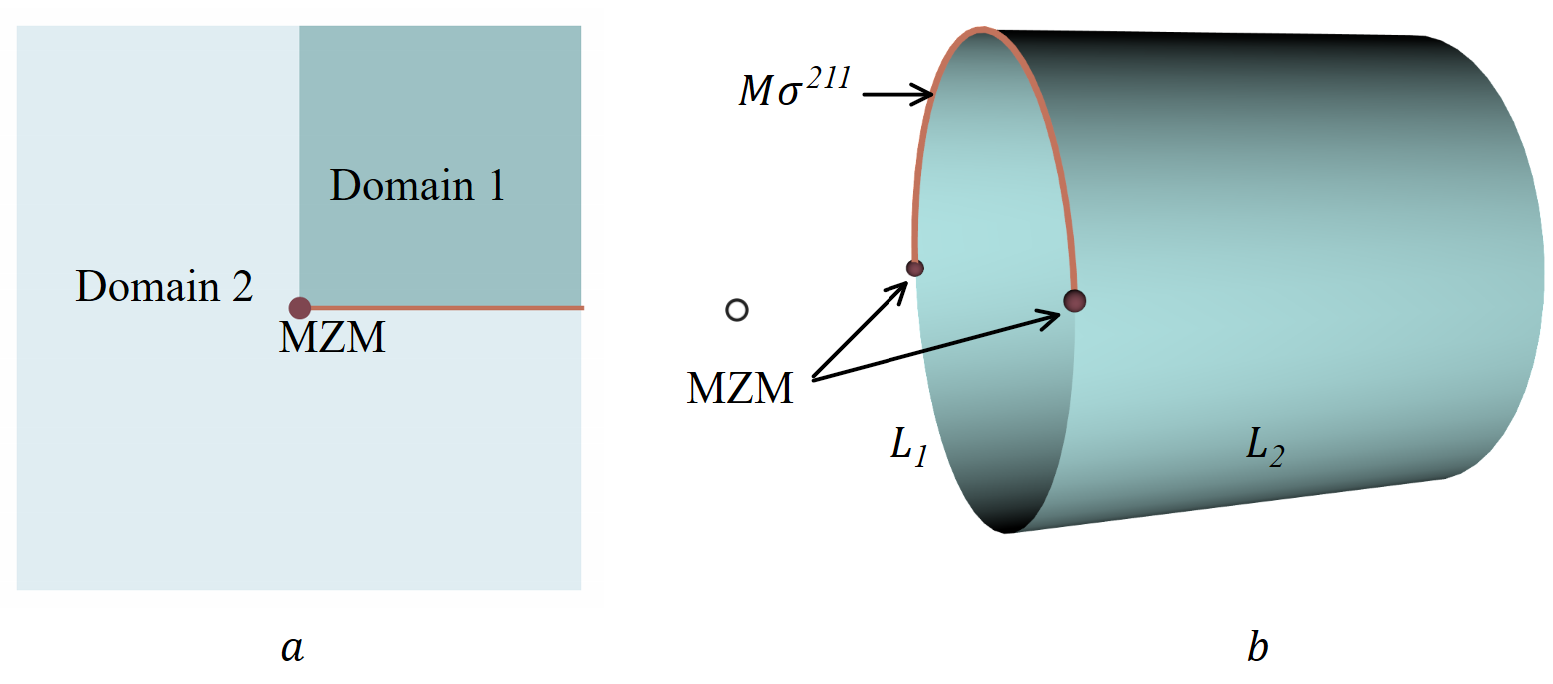}
\caption{$(a)$ The proposed experimental geometry for detecting
the Majorana zero mode (MZM). The entire system is the TDBG in its
spin polarized superconductor phase near half-filling away from
charge neutrality~\cite{kimtalk,TDBG1,TDBG2}. Displacement field
(out of plane) is adjusted such that the valley Chern numbers of
the filled minibands differ by 1 between the two domains. The
vertical domain wall preserves the valley quantum number, while
the horizontal domain wall breaks the valley quantum number
conservation. We propose that a Majorana zero mode (MZM) locate at
the $0d$ junction between the two domain walls. $(b)$ The geometry
of our effective lattice model. The $1d$ domain walls in $(a)$ are
modelled by the open boundary of Eq.~\ref{haldane} defined on a
cylinder with size $20 \times 20$. The horizontal domain wall in
$(a)$ is modelled by an inter-valley scattering $M \sigma^{211}$
on half of the boundary.} \label{fig}
\end{figure}

\section{Topological imprint of VCI: Majorana zero mode}

A deconfined Majorana zero mode (MZM) is a nonabelian anyon
excitation~\cite{wenzero,moorezero,readzero,ivanzero}. MZM can be
engineered in various
designs~\cite{fuzero,zero3,zero4,zero5,zero6}, including the
vortex core of the proximity superconductor at the boundary of a
$3d$ topological insulator~\cite{fuzero2,jiazero}. Later it was
realized that the proximity superconductivity at the boundary of a
$3d$ TI can be directly enforced if the bulk of the system forms a
superconductor when doped with finite charge carrier density. Even
if strictly speaking the bulk superconductor is topologically
trivial, the MZM at the boundary vortex core can still
survive~\cite{ashvinzero}. Recently the topological nature of the
band structure of the iron-pnictides and iron-chalcogenides
materials was
discussed~\cite{ironzero1,ironzero2,ironzero3,irontopo}, and MZM
in the boundary vortex core was observed in iron-based
superconductors~\cite{ironzeroex1,ironzeroex2}.

Here we propose the possibility of a localized MZM at certain
geometric defect of the Moir\'{e} superconductor when the system
is doped away from the VCI. Evidence of spin polarized VCI, and
spin polarized superconductivity were observed in TDBG near
half-filling away from the charge
neutrality~\cite{kimtalk,TDBG1,TDBG2}. We consider the geometry of
Fig.~\ref{fig}$a$, where the displacement field of the two domains
is tuned such that the valley Chern numbers of the filled valence
miniband of the spin-up electron differ by $ \Delta C = 1$ between
the two domains (the spin-down electrons fill both minibands and
have total zero valley Chern number). The vertical domain wall in
Fig.~\ref{fig}$a$ preserves the valley conservation of the
original graphene sheet, while the horizontal domain wall breaks
the valley conservation. Our conclusion is that at the $0d$
junction of the two domain walls, there can be a MZM which can be
viewed as the imprint of the topological effect of the VCI, even
if the system is already doped away from the VCI and forms a
superconductor.

\begin{figure}
\includegraphics[width=0.8\linewidth]{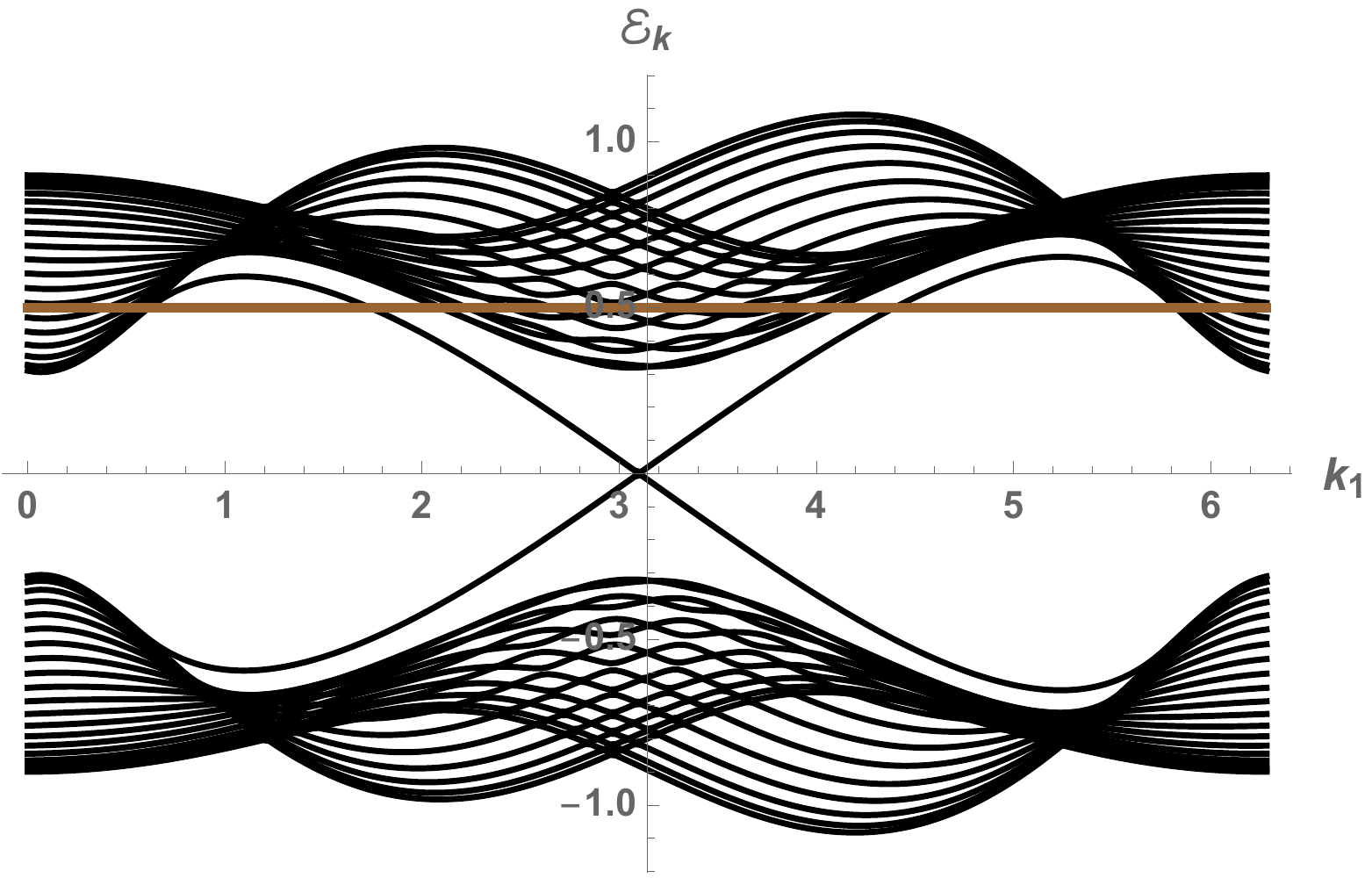}
\caption{The band structure of our effective lattice model for
each valley with two open boundaries. As we can see there are two
counter propagating edge modes from the two boundaries. We have
chosen $t_1 = 0.3$, $t_2 = 0.2$, and $m = 0.05$ in our effective
model Eq.~\ref{haldane}. The horizontal line is the energy
(chemical potential) $\mu = 0.5$. } \label{band}
\end{figure}

\begin{figure}
\includegraphics[width=0.8\linewidth]{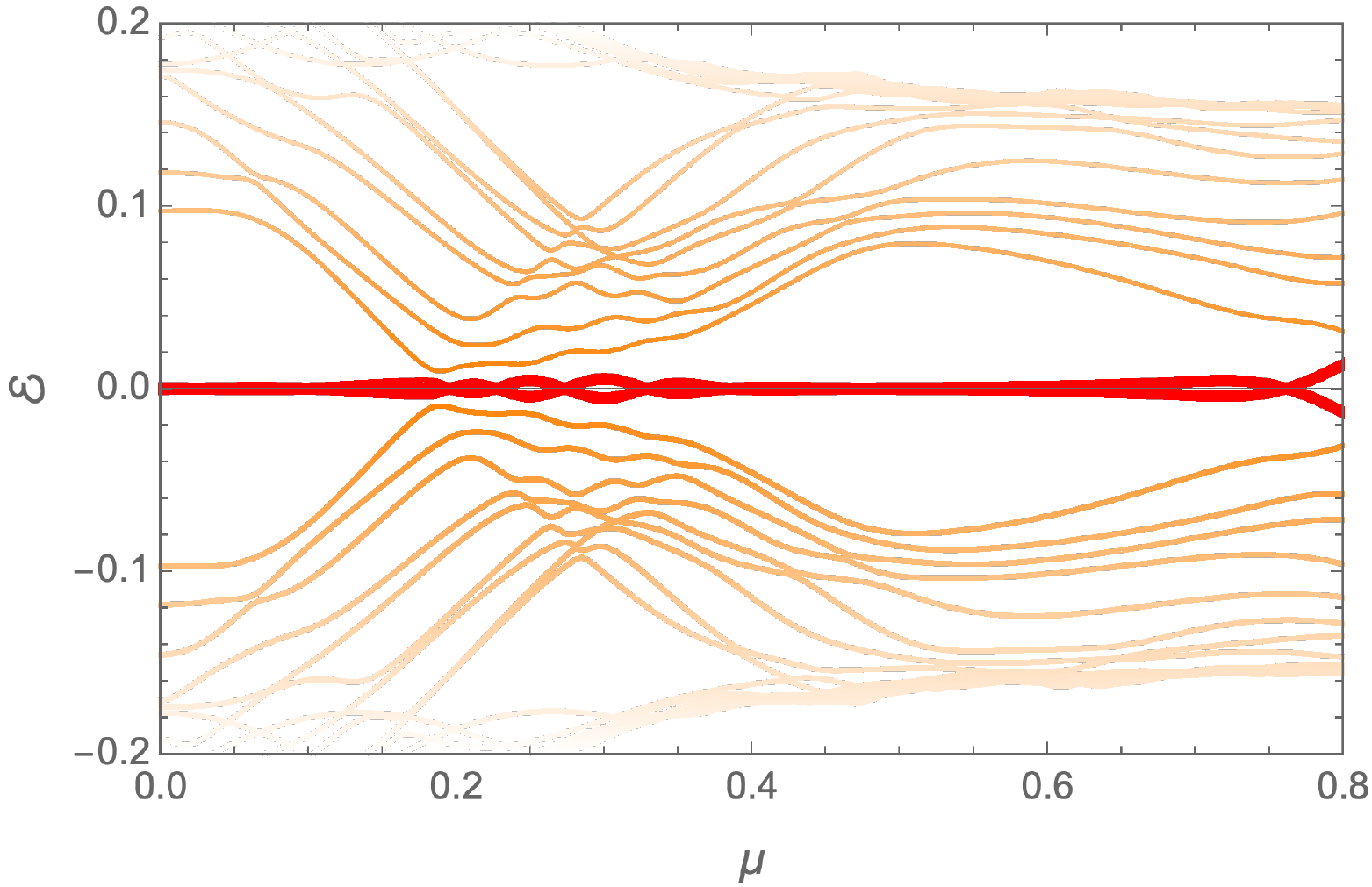}
\caption{The low energy modes of our effective lattice model
plotted against chemical potential, with $s-$wave valley-singlet
pairing amplitude $\Delta = 0.2$ in the bulk, and inter-valley
scattering $M = 0.8$ on half of the boundary. We observe two MZMs
on the two $0d$ junctions between $M$ and $\Delta$ at the
boundary, for a broad range of finite chemical potential, even
when the bulk has finite charge carrier density.} \label{scan}
\end{figure}

\begin{figure}
\includegraphics[width=\linewidth]{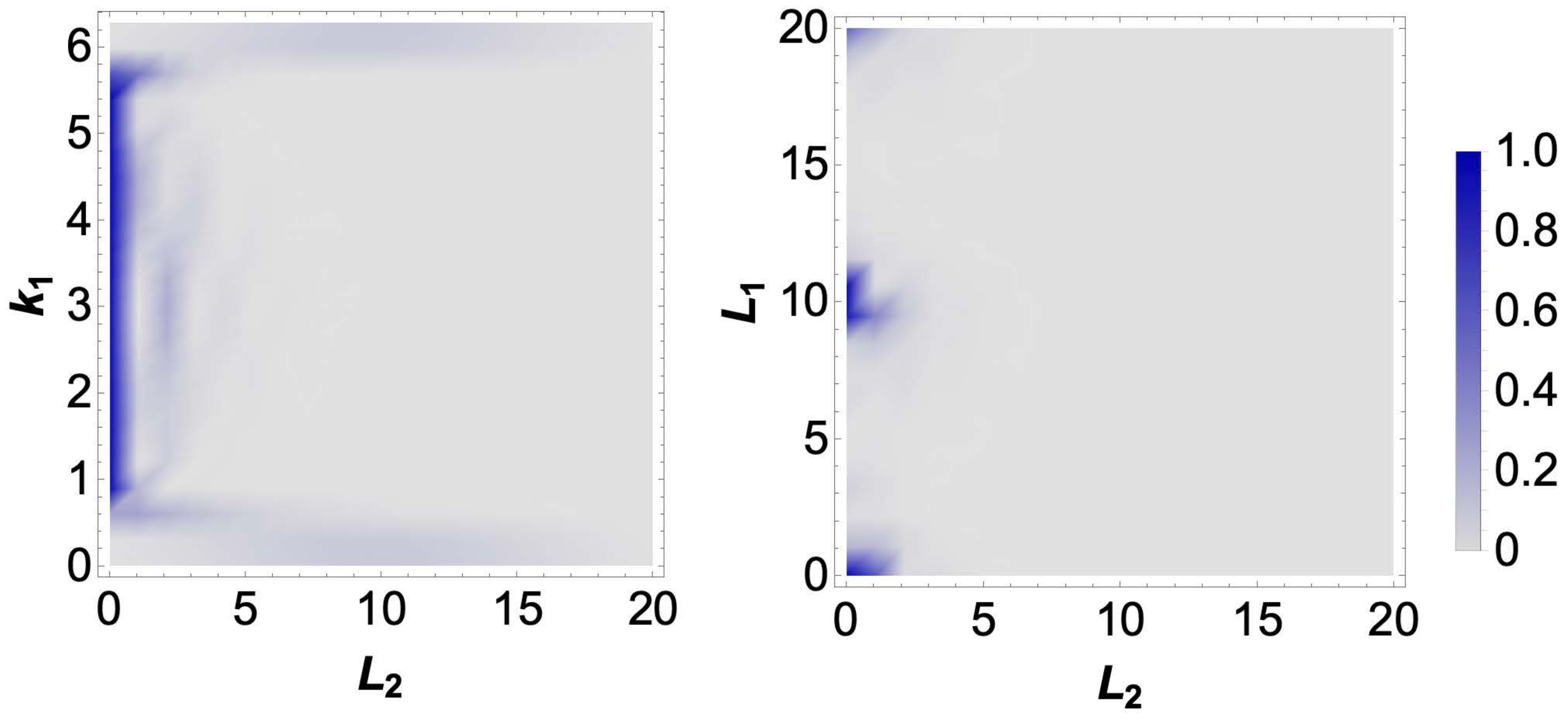}
\caption{$(a)$ the penetration of the spatial wave function of the
edge states (Fig.~\ref{band}) into the bulk of Eq.~\ref{haldane},
plotted against momentum $k_1$ along the boundary; $(b)$ the
spatial wave function of MZMs at the junctions between $\Delta$
and $M$ at the open boundary of our model (Fig.~\ref{fig}$b$),
with chemical potential $\mu = 0.5$, $\Delta = 0.2$, and $M =
0.8$.} \label{wf}
\end{figure}

We model the $1d$ domain wall of Fig.~\ref{fig}$a$ as the open
boundary of an effective model. The contribution from each valley
is modelled by a Haldane's model~\cite{haldanechern} for Chern
insulator defined on a honeycomb lattice: \beqn H_v =
t_{1}\sum_{\left\langle \boldsymbol{i},\boldsymbol{j}\right\rangle
}c_{\boldsymbol{i}}^{\dagger}c_{\boldsymbol{j}} + \eta_v
t_{2}\sum_{\left\langle \left\langle
\boldsymbol{i},\boldsymbol{j}\right\rangle\right\rangle}\ii
c_{\boldsymbol{i}}^{\dagger}c_{\boldsymbol{j}} +
\textrm{H.c.},\label{haldane} \eeqn where $\eta_v = \pm 1$ for the
two different valleys, hence the minibands from the two valleys
carry opposite Chern number. We also turn on a small staggered
potential energy $m \eta_v (-1)^{\boldsymbol{j}}
c^\dagger_{\boldsymbol{j}} c_{\boldsymbol{j}}$ between the two
sublattices to break the symmetry of the system down to $C_3$
(same as the miniband at each valley of the Moir\'{e} system),
while keeping the nontrivial band topology. Please notice that the
lattice of this effective model is not the original graphene
sheet. The unit cell of this effective lattice coincides with the
Moir\'{e} unit cell. The band structure of this model with two
open boundaries (Fig.~\ref{fig}$b$) is plotted in Fig.~\ref{band}.

We turn on a nonzero chemical potential $\mu$, and also a uniform
$s-$wave valley-singlet superconductor order parameter $\Delta$ in
the entire system. The entire bulk Hamiltonian in the Majorana
fermion basis reads: \beqn
\mathcal{H}_{\boldsymbol{k}}^{\textrm{BdG}} &=&
d_{1}\left(\boldsymbol{k}\right)\sigma^{201} +
d_{2}\left(\boldsymbol{k}\right)\sigma^{202} +
d_{3}\left(\boldsymbol{k}\right)\sigma^{033} \cr\cr &-&
\mu\sigma^{200} + \Delta\sigma^{120} + m \sigma^{233}, \cr\cr
d_1(\boldsymbol{k}) &=& t_1(\cos k_1 + \cos k_2 + 1), \cr\cr
d_2(\boldsymbol{k}) &=& t_1 (\sin k_1 + \sin k_2) , \cr\cr
d_3(\boldsymbol{k}) &=& 2 t_2(\sin k_2 - \sin k_1 + \sin(k_1 -
k_2)), \label{eq:Haldane BdG} \eeqn where $k_1 = k_x$, $k_2 =
\frac{1}{2}k_x + \frac{\sqrt{3}}{2}k_y$. $\sigma^{abc} =
\sigma^{a} \otimes \sigma^b \otimes \sigma^c$, the three Pauli
spaces correspond to the (1) real and imaginary parts of the
electron operator, (2) the valley space, and (3) the sublattice
space of the effective Haldane's model respectively.

The horizontal domain wall in Fig.~\ref{fig}$a$ breaks the valley
conservation, and hence will induce inter-valley scattering
between the edge states of the two valleys. This effect is
modelled by an inter-valley scattering $M\sigma^{211}$ on half of
an open boundary (Fig.~\ref{fig}$b$). We expect one MZM on each of
the $0d$ junction between $\Delta$ and $M$ on the $1d$ boundary.
With zero chemical potential, the existence of the MZM can be
straightforwardly inferred from the Jackiw-Rebbi
solution~\cite{jackiwzero}. In fact, since the bulk is always
gapped, we can view the $1d$ segment with nonzero backscattering
$M$ as a $1d$ Kitaev's chain. At $\mu = 0$, this Kitaev's chain is
in its topological nontrivial phase based on the Jackiw-Rebbi
solution, and there is a MZM at its end. The Kitaev's chain can
still be in the topological nontrivial phase even when the bulk
has finite charge density, since the bulk is gapped due to
superconductivity, hence the MZMs can still persist. Indeed, we
observe two MZMs with a broad range of chemical potential even
when the bulk band has nonzero charge density (Fig.~\ref{scan}).
The spatial wave function of these MZMs are plotted in
Fig.~\ref{wf} with chemical potential $\mu = 0.5$ (when the bulk
has finite charge density), pairing amplitude $\Delta = 0.2$, and
inter-valley scattering $M = 0.8$ in our effective model.

Here we stress that a single localized Majorana fermion zero mode
is stable against any perturbation, regardless of the symmetry of
the system. This is why the classification of the Kitaev's
chain~\cite{kitaevclass,ludwigclass1,ludwigclass2} without any
symmetry is $\mathbb{Z}_2$, $i.e.$ there is nontrivial topological
superconductor in $1d$ whose edge state is a single Majorana
fermion zero mode which is stable against perturbations without
assuming any symmetry. Multiple MZMs can be created by designing
more corners of the domain wall in Fig.~\ref{fig}$a$. Effective
braiding of these MZMs can be achieved, and their nonabelian
statistics can be revealed by ``measuring" these MZMs
consecutively, without actually braiding them
spatially~\cite{nayakmeasure1,nayakmeasure2}.

\section{Discussion}

In this work we have discussed the interacting valley Chern
insulator (VCI) that can be realized in Moir\'{e} systems, and its
classification under interaction. We demonstrate that although
without interaction the classification of the VCI is $\mathbb{Z}$,
interaction reduces the classification down to $\mathbb{Z}_3$,
hence a VCI with valley Chern number $3$ becomes a trivial
insulator under interaction. This result implies that the VCI with
valley Chern number 3 can be deformed continuously into a product
state (atomic insulator) under interaction that preserves all the
symmetries.

It is well-known that a localized Wannier orbital cannot be
constructed using single particle states in a band with nontrivial
band topology, which causes obstacles of using the standard
tight-binding model and Hubbard model to describe the strong
interaction effect when the topological band is doped with charge
carriers. But since we demonstrated that VCI with valley Chern
number 3 is trivialized by interaction, a new simple formalism to
study doped interacting VCI with valley Chern number 3 becomes
possible. In particular, it may be possible to construct an
``interacting local Wannier orbital" which is not a simple linear
combination of single particle operators, but a combination of
fermionic composite operators. This observation, if successful,
will simplify theoretical study of some of the Moir\'{e} systems
tremendously. We will leave this observation to future
explorations.

The mapping of interacting VCI to a bSPT also has potentially
interesting consequences. The phase transition between a VCI with
valley Chern number 1 and a trivial insulator is described by two
massless Dirac fermions, and the extra time-reversal symmetry
$\mathcal{T}$ mentioned in the text will guarantee one direct
transition. While the topological-to-trivial phase transition of a
bSPT state is described by a $N_f = 2$ QED$_3$ with noncompact
$\U(1)$ gauge field~\cite{xudual,deconfinedual}. The topological
equivalence between the interacting VCI and bSPT implies that the
physics around that topological-to-trivial phase transition of the
interacting VCI can have two very different descriptions that are
based on different quantum field theories.

{\it --- Acknowledgement}

Chao-Ming Jian is supported by the Gordon and Betty Moore
Foundations EPiQS Initiative through Grant GBMF4304. Cenke Xu is
supported by NSF Grant No. DMR-1920434, and the David and Lucile
Packard Foundation. The authors thank Chetan Nayak for helpful
discussions, and thank Ashvin Vishwanath for pointing out
Ref.~\onlinecite{ashvinzn}.

\bibliography{IVCI}

\end{document}